%Plain TeX file using harvmac and epsf
%A proof of the irreversibility of renormalization group flows
%by S Forte and JI Latorre

\input harvmac
\input epsf
%\draftmode
%\magnification\magstep1
\parskip 6pt
\def\sqr#1#2{{\vcenter{\hrule height.#2pt \hbox{\vrule width.#2pt
height#1pt \kern#1pt \vrule width.#2pt} \hrule height.#2pt}}}
\def\sq{{\mathchoice\sqr55\sqr55\sqr{2.1}3\sqr{1.5}3}\hskip 1.5pt}
\def \l{\langle}
\def \r{\rangle}

\def \pr{\partial}

\def \d{{\rm d}}

\def \darrow#1{\raise1.5ex\hbox{$\leftarrow$}\mkern-16.5mu #1}
\def\darr#1{\raise1.5ex\hbox{$\leftrightarrow$}\mkern-16.5mu #1}

%References
\lref\thooft{G. 't Hooft, {\sl Recent developments in 
gauge theories}, Eds. G. `t Hooft {\sl et al.}, 
Plenum Press, New York, 1980.}
\lref\adler{S. L. Adler and W. Bardeen, Phys. Rev. {\bf 182} (1969) 1517.}
\lref\wilson{K. G. Wilson and J. Kogut, Phys. Rep. {\bf 12C} (1974) 75.}
\lref\wallace{D. J. Wallace and R. K. P. Zia, Ann. of Phys.
{\bf 92} (1975) 142.}
\lref\zamref{A. B. Zamolodchikov, JETP Lett. {\bf 43} (1986)
730.}
\lref\burgess{C.~J.~C.~Burges {\sl et al.},
 Ann. Phys. {\bf 167} (1986) 285.}
\lref\cardy{J. L. Cardy, Phys. Lett. {\bf B215} (1988) 749.}
\lref\osbornone{H. Osborn, Phys. Lett. {\bf B222} (1989) 97\semi
    I. Jack and H. Osborn, Nucl. Phys. {\bf 343} (1990) 647.}
\lref\cfl{A. Cappelli, D. Friedan  and J.I. Latorre,
    Nucl. Phys. {\bf B352} (1991) 616.}
\lref\cfv{A. Cappelli, J. I. Latorre and X. Vilas\'\i s-Cardona,
    Nucl. Phys. {\bf B376} (1992) 510; hep-th/9109041.}
\lref\flv{D. Z. Freedman, J. I. Latorre and X. Vilas\'\i s-Cardona,
    Mod. Phys. Lett. {\bf A6} (1991) 531.}
\lref\zumbach{G. Zumbach, Nucl.Phys. {\bf B413} (1994) 754; Phys. Lett.
{\bf A190} (1994) 225.\semi
   J. Generowicz, C. Harvey-Fros, T. R. Morris,
Phys. Lett. {\bf B407} (1997) 27; hep-th/9705088\semi
P. Haagensen {\sl et al.}, Phys. Lett. {\bf B323} (1994) 330;
hep-th/9310032.}
\lref\fradkin{A.H. Castro Neto and E. Fradkin,
Nucl. Phys. {\bf B400} (1993) 525.}
\lref\bastianelli{F. Bastianelli, Phys. Lett. {\bf B369} (1996) 249-254;
hep-th/9511065.}
\lref\freedman{D. Anselmi, J. Erlich, D. Z. Freedman, A. Johansen;
hep-th/9711035\semi
D. Anselmi, D. Z. Freedman, M. T. Grisaru, A. A. Johansen;
hep-th/9708042.}
\lref\shore{G. M. Shore, Phys. Lett. {\bf B253}
(1991) 380; {\bf B256} (1991) 407.}
\lref\polyakov{A. Polyakov, Phys. Lett. {\bf B103} (1981) 207.}
\lref\wirev{
C. Callan, S. Coleman and R. Jackiw, Ann. of Phys. {\bf 59}
(1970) 42\semi
S. Coleman, {\it Aspects of Symmetry},
(Cambridge, Cambridge, U.K., 1985)\semi
R.~Jackiw, in S.~B.~Treiman {\sl et al.}, 
{\it Current Algebra and Anomalies} (World Scientific,
Singapore, 1985).} 
\lref\widet{J. C. Collins, {\it
Renormalization}, (Cambridge, Cambridge, U.K., 1984).}
\lref\scwir{J. C. Collins, A. Duncan and S. D. Joglekar, {\it
Phys. Rev.} {\bf D16} (1977) 438\semi N. K. Nielsen, {\it Nucl. Phys.}
{\bf B120} (1977) 212\semi R. Tarrach, {\it Nucl. Phys.} {\bf B196}
(1982) 45.}
\lref\osborntwo{H. Osborn and S. Petkou
    Ann. Phys. {\bf 231} (1994) 311; hep-th/9307010.}
\lref\osbornthree{J. I. Latorre and H. Osborn, Nucl. Phys. 
{\bf B511} (1998) 737; hep-th/9703196.}
\lref\osbornfreedman{H. Osborn and D. Z. Freedman, hep-th/9804101.}
\lref\dusedau{D. W.  D\"usedau and D. Z. Freedman, Phys. Rev. {\bf D33}
(1986) 395.}
\lref\camporesi{R. Camporesi, Phys. Rev. {\bf D43} (1991) 3958\semi
R. Camporesi and A. Higuchi, Phys. Rev. {\bf D45} (1992) 3591.}
\lref\vilenki{N. Ja. Vilenkin, {\sl Special functions and 
the theory of group representations},  American Math. Soc. 1968.}
\lref\dr{D. Z. Freedman, K. Johnson and J.I. Latorre,
    Nucl. Phys. {\bf B371} (1992) 353.}
\lref\duff{M. J. Duff, Nucl. Phys. {\bf B125} (1977) 334.}
\lref\birrell{N. D. Birrell and P. C. W. Davies, 
{\sl Quantum fields in curved space}, Cambridge Univ. Press (1982).}
\lref\unpublished{A. Cappelli, D. Friedan and J. I. Latorre, 
unpublished work.}
\noblackbox
\pageno=0\nopagenumbers\tolerance=10000\hfuzz=5pt
\baselineskip=12pt plus2pt minus2pt
\newskip\footskip\footskip=10pt 
\rightline{hep-th/9805015}
\rightline{UB-ECM-PF 98/12}
\rightline{DFTT 17/98}
\vskip 10pt
\centerline {\bf A PROOF OF THE IRREVERSIBILITY }
\centerline {\bf OF RENORMALIZATION
GROUP FLOWS  IN FOUR DIMENSIONS}
\vskip 30pt
\centerline {Stefano Forte$^{(a, c)}$ and  Jos\'e I. Latorre$^{(a, b)}$}
\vskip 20pt
\centerline{${}^{(a)}${\it
Departament d'Estructura i Constituents de la Mat\`eria,}}
\centerline{\it Universitat de Barcelona}
\centerline{and}
\centerline{ ${}^{(b)}${\it I.F.A.E.,}}
\centerline {\it Diagonal 647, E-08028 Barcelona, Spain}
\vskip 10pt
\centerline {${}^{(c)}${\it INFN, Sezione di Torino},}
\centerline {\it Via P. Giuria 1, I-10125 Torino, Italy}
\vskip 40pt

{\centerline{\bf Abstract}
\medskip\narrower
\baselineskip=10pt

We present a proof of  the irreversibility of
renormalization group flows,  i.e. the
 $c$--theorem
for unitary, renormalizable theories in four 
(or generally even) dimensions.
Using Ward identities for scale transformations and
spectral representation arguments,  we show that the 
$c$--function based on the trace of the energy-momentum tensor
(originally suggested by Cardy) decreases
monotonically along renormalization group trajectories.
At fixed points this $c$--function is stationary and coincides with
the coefficient of the Euler density in the trace anomaly, while
away from fixed points its decrease is due to the decoupling
of positive--norm massive modes.\smallskip}
\vskip 48pt

\centerline{Submitted to: {\it Nuclear Physics B}}
\vskip 64pt
\line{April 1998\hfill}

\vfill\eject \footline={\hss\tenrm\folio\hss}

\newsec{Zamolodchikov's two dimensional $c$--theorem}

  Fundamental theorems that impose constraints on the long-distance
realization of a quantum field theory are rare. An instance of
such a powerful, non-perturbative tool is provided by
the t'Hooft anomaly matching  conditions~\thooft. If
an ultraviolet (UV) theory has an
axial anomaly, as a consequence of the Adler-Bardeen non-renomalization 
theorem~\adler\ the anomaly coefficient must be matched by that of
 its infrared (IR)
effective realization.
This exact result is useful to check the mutual consistency of short {\it vs.}
long distance realizations of a given theory.
Yet, the same non-renormalization theorem
which provides the  basis of anomaly matching trivializes the relation
between different scales and thus the way the IR limit is attained. 
This motivates the search
for other theorems which may add further and more general
constraints to the IR realization of quantum field theories.

  The mechanism controlling the modification of
a physical theory through a change of scale is the renormalization
group (RG). Following Wilson \wilson, we may set up an exact RG equation
which describes the change of the effective hamiltonian of
a theory as we flow towards the IR. The derivation of
this equation proceeds in two steps: an
 integration of modes
(Kadanoff block-spin transformation) followed by a rescaling.
Since integrating out modes seems an irreversible operation, 
 one is naturally lead to
ask whether the RG flow itself is irreversible. This is equivalent to
asking  whether there is a fundamental
obstruction to recover microscopic physics from
macroscopic physics, or, more generally, whether there is a
net information loss along RG trajectories.
Indeed, realistic
quantum field theories seem intuitively irreversible; on the other
hand, however, it is known that 
 some theories may display limit cycles, and thus their  RG flow  is
manifestly not irreversible.
The question is, thus, to see under
which conditions  irreversibility may
follow.

Efforts to answer this question started with
 the work of ref.~\wallace, where
the perturbative
expansion of a theory of $N$ scalars with interaction
$g_{abcd}\phi^a
\phi^b \phi^c \phi^d$ ($a,b,c,d=1\dots N$) 
is analyzed up to three loops. The beta functions of the theory turn
out to be gradients of a scalar function of the couplings,
thus yielding irreversible flows at this order of perturbation theory.

A fundamental theorem was later proven by Zamolodchikov \zamref\
in the context of two dimensional field theories. This result is based on the
construction of a function of the couplings $c(g^i)$  from two-point
correlators of energy-momentum tensors,
 which is then proven
to decrease monotonically
along RG trajectories
for unitary theories. More precisely, defining
at the operator level a theory as
\eqn\theory{S=g^i \int_x{\cal O}_i(x)\ ,}
one can construct a function $c(g^i)$ whose RG flow is
\eqn\zam{-\beta^i {\pr\over \pr g^i} c(g^j) = - \beta^i\beta^j
G_{ij}
 \leq 0\ , }
where $G_{ij}$ (known as the
Zamolodchikov metric) is given by
\eqn\zammetric{G_{ij}\equiv \left.\l{\cal O}_i(x) {\cal O}_j(0)\r
\right|_{x=\mu^{-1}}\geq 0\qquad \mu={\rm reference \ scale}\ ,}
and  is  positive definite due to reflection positivity.
 Because at fixed points the function $c(g^i)$
reduces to  the central
charge of the conformal field theory which defines it,
this remarkable result is known as  the $c$--theorem. It can 
furthermore  be proven that the beta functions are gradients
of $c(g^i)$ to the first few orders of conformal perturbation theory, namely,
in a perturbative expansion around any (gaussian or non-gaussian)
fixed point.

The result of ref.~\zamref\ shows that the sought--for  necessary condition
for irreversibility is unitarity. The proof of the theorem
makes also use of renormalizability and Poincar\' e invariance {\sl via}
the absence of anomalous dimensions
for the energy momentum tensor and  its conservation. The construction
of a natural positive definite metric $G_{ij}$ in the space
of couplings
 introduces the concept of distance between theories, each
defined at a given scale.
 The difference of the $c$--charges between two fixed points
is reparametrization invariant.
  RG equations are simply equations
of motion in the space of theories.

The essence of the $c$--theorem in two dimensions can thus be
summarized as follows.
For every unitary, renormalizable, Poincar\'e invariant quantum field
theory there exists a universal $c$--function which decreases 
along RG flows, while it is only stationary at (conformal) fixed points, where
it reduces
to the central charge. This sets an arrow on RG flows, and implies their
irreversibility. It follows that a theory can be the
IR realization of a given UV theory only if their
central charges satisfy the inequality $c_{IR} < c_{UV}$.
It is important to notice that the identification of the $c$--function
with the central charge at conformal points makes its determination
possible and thus the theorem useful for realistic applications.

One may thus conjecture that  irreversibility may also hold in four
and higher dimensions. Evidence supporting
this conjecture is provided by 
perturbative computations \cfv,
by the non-trivial leading order of the derivative expansion 
of the exact RG equation \zumbach, and by 
a large number of
explicit non-trivial RG flows~\refs{\bastianelli,\freedman}. 
However, the generalization of the $c$--theorem itself
to higher dimensional field
theories has proven to be problematic \refs{\cardy\osbornone-\cfl}.
It is the purpose of this
paper to present a proof of irreversibility of RG group flows 
in four and higher
dimensions. This proof will
include and synthesize ideas  which originate from various lines of
research. First and more importantly, we will make use of a
four-dimensional generalization of the $c$--function proposed by 
Cardy \cardy. 
Furthermore, our construction will make use of the physical
insight provided by the alternative proof of the 
two-dimensional $c$--theorem of
ref.~\cfl,  based on the
spectral analysis of two-point energy-momentum tensor
correlators.
Combining this with the use of scale Ward identities
will allow us to prove that Cardy's $c$--function decreases along RG
flows, thus establishing the $c$--theorem in four dimensions,
and generally even dimensions.

\bigskip\bigskip

\newsec{Spectral proof of Zamolodchikov's theorem}

The meaning of the two--dimensional $c$--theorem can be understood in
a somewhat more general way by means of techniques first introduced
in ref.~\cfl, which we now briefly review.
The correlator of two energy-momentum
tensors can be written using a spectral representation as
\eqn\spectral{\l T_{\alpha\beta}(x) T_{\mu\nu}(0)\r =
{\pi\over 3}  \int \d\lambda\ \rho(\lambda,\mu)
\left(\pr_\alpha\pr_\beta - g_{\alpha\beta} \sq\right)
\left(\pr_\mu\pr_\nu - g_{\mu\nu} \sq\right)
G(x,\lambda)\ ,}
where $\lambda$ is the spectral parameter (with dimensions of mass),
$\rho(\lambda,\mu)$ is the spectral function, which depends on
$\lambda$ and on the subtraction point $\mu$,
and $G(x,\lambda)$ is the free scalar propagator of a particle with mass
$\lambda$.

At a fixed point, the spectral function reduces to a delta function 
\eqn\fixedpoints{\left. \rho(\lambda,\mu)\right|_{cft}=
c\ \delta(\lambda)\ ,}
where the coefficient $c$, being dimensionless, cannot depend on $\mu$ on
dimensional grounds, and is thus
necessarily a  constant. This reflects the fact that all
physical intermediate states in \spectral\ are massless.
At an UV fixed point we thus get that 
\eqn\cuv{c_{UV}=\int \d\lambda\ \rho(\lambda,\mu)\ .}
By considering $\rho(\lambda,\mu)$ in the vicinity of an UV fixed point it
can also be shown explicitly that $c$ in eq.~\cuv\ coincides  with the central
charge of the conformal field theory.
Now, the integral of the spectral density cannot depend on $\mu$,
again on dimensional grounds, so that eq.~\cuv\ actually holds
for all finite $\mu$ (i.e. away from the IR limit $\mu\to\infty$).

On the other hand, in the IR limit only massless modes survive, so that
the spectral function can in general be written as
\eqn\formofc{\rho(\lambda,\mu)= c_{IR}\ \delta(\lambda)
 + \rho_{\rm smooth}(\lambda,\mu)\ ,}
where the contribution of all massive modes is contained in $\rho_{\rm
smooth}(\lambda,\mu)$. 
It thus follows that
\eqn\inequal{c_{UV}=c_{IR}+\int \!\d\lambda\, \rho_{\rm smooth}(\lambda,\mu)\
,}
where the second term on the r.h.s. is necessarily $\mu$--independent.
Notice that this means that the IR limit is not uniform, i.e. it does not
commute with the integration over $\lambda$, so that the IR limit of
$c$ does not coincide with the constant value  $c_{UV}$.
Finally,  unitarity guarantees that $\rho_{smooth}$ is positive, so
\eqn\finaldeltac{c_{UV}-c_{IR} \geq 0\ .}
Several examples illustrating the above construction can be found in
ref. \flv.

This alternative proof of the
$c$--theorem provides us with a number of physical insights. 
First, we see that unitarity is necessary for irreversibility because 
it guarantees positivity of the spectral
representation. Furthermore, we understand that
the decrease of the $c$--function is
due to the decoupling of massive modes in the spectrum which
appear as intermediate states in the spectral decomposition.
Finally, we see that the central charge  provides an
effective counting of massless degrees of freedom, and must thus
be an additive function:
RG flows are then such that the effective counting of massless
modes in the IR is lower than in the UV.
The physical picture is thus the following. 
At the UV  fixed point, all modes
are massless and appear as a delta term in $\rho$. When the scale
parameter $\mu$ starts to move, only some modes remain massless and
form $c_{IR}$, the rest acquire a mass and start to decouple. In
the IR theory only the $c_{IR}$ contribution remains.

The original proof of ref.~\zamref\ is based on the construction of 
a particular combination of
correlators, which is then shown to  decrease along RG flows. It is
now clear, however, that the $c$--theorem can be formulated at the level
of the density of states.
One may then construct many  combinations
of correlators to prove the theorem, but all of them are related to
a unique spectral density.    

A technical but important point concerning this proof is related
to  subtractions in the two-point correlator \spectral.
The variation  $\Delta c=c_{UV}-c_{IR}$ is
given by the integral of $\rho_{\rm smooth}$ which is free of
subtraction ambiguities, as it picks up the
non-local contribution of   intermediate massive
modes. In other words, $\Delta c$ can be computed from the
 imaginary part of $\l T T\r$, which is free of contact terms and,
thus,
scheme independent. However, the correlator itself is only defined up
to contact terms, which can be freely changed by choice of renormalization
scheme. This freedom is  restricted by the 
fact that the energy-momentum tensor
obeys an algebra which fixes the relative subtractions in one-- and
two--point functions. This point will be of great relevance in the
proof of the $c$--theorem in four
dimensions.

A first important obstacle in trying to generalize the
spectral arguments
to higher dimensions is due to the fact that the proof  of
ref.~\zamref\  makes
essential use of a peculiarity of two dimensions, namely the fact
that there is no spin. It follows that  
correlators of two  energy-momentum tensors
are saturated by spinless intermediate states, i.e., the response to
scale  and shear transformations are controlled by one single
spin structure. In four (and higher) dimensions,
we find two separate structures in the spectral decomposition:
one  corresponds
to spin--zero and the other to spin--two
intermediate states. A number of branching avenues thus
opens as one tries  to define a candidate for a $c$--function
\refs{\osbornone,\cfl,\unpublished,\shore}. 
One may in particular conjecture~\cfv\ that 
the natural generalization of the theorem should lie in the
spin--zero part of the spectral decomposition of the
   $\l T T\r$ correlator, on the grounds that this is related to the
trace of the energy-momentum tensor and thus to the response of the
system upon dilatations.
However, dimensional analysis
shows
that in $d$ dimensions
\eqn\problem{\left.\rho^{spin=0}(\lambda,t)\right|_{cft}= c\ \lambda^{d-2}
\delta(\lambda)\ ,}
that is,
the would-be analogue of $c$ in eq.\fixedpoints\
is now inaccessible to physical observables
as it multiplies a vanishing distribution for $d > 2$.

\bigskip\bigskip

\newsec{Cardy's proposal for a $c$--charge}

A different avenue towards the derivation of a higher--dimensional
generalization of the $c$--theorem is based on reconsidering the role
of the central charge in conformal field theories which underlies the
two--dimensional proof. 
Recall that the central charge
 of a conformal field theory appears as the
coefficient of the identity operator in the operator
product expansion (OPE) of two energy-momentum tensors
\eqn\opett{T(z) T(0) \sim {c\over 2 z^4} I +\dots \ .}
and, consequently, in the numerator of the $\l T(z) T(0)\r$ correlator.
Now,  $c$ can also be understood \polyakov\ as the
coefficient of the trace anomaly of a conformal field
theory in curved space
\eqn\canomaly{\l T^\alpha_\alpha\r_{cft}
\equiv \l\theta\r_{cft}= -{c\over 12}  R\ .}
Moreover, the integrated trace anomaly turns out to be proportional
to the Euler number, $\chi(M)$, of the background manifold $M$
\eqn\intanomaly{\int_M \sqrt{g}
\l\theta\r_{cft} = -{c\over 12} \int_M \sqrt{g} R =
- c{\pi\over 3}\ \chi(M)\ .}
The central charge can thus also be viewed as the coefficient of a
contribution to
the trace anomaly which admits a topological interpretation.

This different understanding
of the central charge suggests a new generalization to higher 
dimensions~\cardy:
At fixed points, the $c$--function should reduce to a coefficient
hidden  in the trace anomaly. In four dimensions, we
explicitly have \refs{\duff,\birrell}
\eqn\tracefour{\l \theta(x)\r_{cft} ={1\over 2880} \left(
-3 c_F F(x) + c_G G(x) + c_R \sq R(x)\right)}
where
\eqn\fandg{\eqalign{&F=C^2_{\mu\nu\rho\sigma}=
R^2_{\mu\nu\rho\sigma}-2 R^2_{\mu\nu}+{1\over 3} R^2\cr
&G=R^2_{\mu\nu\rho\sigma}-4 R^2_{\mu\nu}+ R^2 \ .\cr}}
The coefficients $c_F$  and $c_G$ seem  {\sl a priori} equally good
candidates for the $c$--functions, while $c_R$ isn't,
since it multiplies a divergence operator and it is thus scheme dependent.
The coefficient $c_F$, however, which can be shown \cfl\ to
coincide  with the positive coefficient of the spin--two structure in
the correlators of two energy-momentum tensors, is found not to 
decrease 
along RG flows in specific examples, 
and is ruled out.

The only remaining possibility is thus $c_G$. This
can be singled out by taking an integral over {\sl e.g.} a sphere,
\eqn\centralb{\l \int_{S^n} \sqrt{g} \theta\r_{cft}= {\pi^2\over 90}
c_G\,\chi(S^n)\ ,}
where
 $\chi(M)=\int_{S^n} \sqrt{g}\, G$ is the Euler characteristic of
the manifold. The $c_R$ contribution then vanishes
because it is a total derivative, while the $c_F$ contribution
vanishes because it is 
proportional to the square of the Weyl tensor and
the sphere is conformally flat. It is thus natural to propose \cardy\ 
to consider the integrated
trace anomaly on a sphere
\eqn\cardycandidate{c=\int_{S^n} \sqrt{g} \l \theta \r}
as a candidate for the $c$--function in higher (even) dimensions.
Notice, however, that away from conformal points 
a variety of different choices for the $c$--function is possible, 
all of which reduce to the same $c$--charge eq.~\centralb\ at
conformal points: namely, any combination of the charge of
eq.~\centralb\  and
terms proportional to the beta--functions of the theory. 
In particular, the choice eq.~\cardycandidate\ 
itself, away from conformal points, does not reduce to the charge of
eq.~\centralb, since the trace of the 
energy momentum tensor then receives extra contributions (proportional
to the beta functions) on top of it,
 due to the breaking (canonical
and anomalous) of dilatation symmetry.

The candidate $c$--function eq.~\cardycandidate\  
can be determined in the first few orders of
conformal  perturbation theory and shown
explicitly to decrease along RG flows~\cardy.
However, in order to establish 
a general theorem one must determine the  variation of eq.~\cardycandidate\
along generic RG trajectories. Naively, this is related
to the correlator of two traces of the energy momentum tensor, but this
contains subtractions which render the proof of positivity non-trivial.

The $c$--function eq.~\cardycandidate\ 
has a number of desirable properties. It is
based on the energy-momentum tensor, which exists for any theory
and couples to all degrees of freedom. It is related to the
integrated
trace of this operator, which generates  scale transformations, and thus
implement  RG flows. 
Finally, it is additive,  as required for a function that should
count effective degrees of freedom.

On the other hand,  the need to resort to a curved background
may seem  unnatural. However, the coefficient $c_G$ is also 
unambiguously determined
as a coefficient in three point functions
of the energy momentum tensor. In general, in $d=2n$
dimensions, anomalies show up in $n+1$--point functions, as well as
 in the one--point trace anomaly in curved space, and the choice of the
latter form is a matter of technical convenience. The background metric
plays here a similar role to that of an 
external current when discussing anomalies
in fermionic loops.

The conjecture that $c$ defined in 
eq.~\cardycandidate\ may be a candidate for the 
$c$--function has been tested in various ways. First, this $c$--function
can be verified to be 
positive in all known conformal field theories. This
fact can be related to a modified weak positivity theorem in
quantum field theory \osbornthree. Furthermore, it can be explicitly
checked that 
the coefficient of the trace anomaly 
obeys the desired inequality eq.~\finaldeltac\ in an impressively large 
 number of non-trivial cases where the UV and IR realizations
are known exactly~\refs{\bastianelli,\freedman}.
Finally, it can be shown 
that no combination  of $c_F$ and  $c_G$ can satisfy the inequality
except $c_G$ itself \freedman. 

A different proposal for
a $c$--function away from conformal field theory, consisting of a
different combination of the charge eq.~\centralb\ and a term
proportional to $\beta^i$, has also been considered~\osbornone\
(see also ref.~\osbornfreedman). Of course, this candidate
$c$--function also passes all
successful tests based on the difference of
UV and IR $c$--charges, which only depend on the value of the function
at conformal points. 
We will not discuss this interesting possibility
 further and concentrate
on Cardy's conjecture eq.~\cardycandidate.

\bigskip\bigskip

\newsec{Proof of irreversibility}

We will now present a proof of
  irreversibility of RG trajectories for the trace of the
energy-momentum tensor in four dimensions. Our proof will
actually remain valid for any even number of dimensions.
We will consider the integrated trace of the energy-momentum tensor
on curved space. For simplicity, we take a
$d$--dimensional maximally symmetric space,
where the energy-momentum tensor satisfies
\eqn\emomsym{\langle T^{\mu \nu}(x)\rangle=\langle\theta\rangle
{g^{\mu\nu}(x)\over d},}
 where $\l\theta\r$ is an $x$--independent constant,
so that  integration of the trace over space amounts to a  volume
factor which can be absorbed in the normalization.
We further take the  curvature to be the
negative constant $R=-a^2 d (d-1)$, where $a$ is a
constant parameter with the dimensions of mass.
None of these choices
entails loss of generality, even though in the generic non-compact
case, where the volume is infinite, the space
integration has to be defined more carefully.

We then define
a dimensionless $c$-function depending on the subtraction point
 $\mu$ and 
the renormalized couplings $g^i$ as 
\eqn\candidate{c\left({\mu\over a},g^i\right)
 \equiv {a^{-d} V\over A_d} \l\theta\r\ ,}
where $V$ is the volume of the $(d-1)$--dimensional sphere
and $A_d$
is a normalization factor which we fix as
\eqn\norm{
V={2 \pi^{d\over 2}\over \Gamma\left({d\over 2}\right)} \qquad,\qquad 
A_d=B_d \left( d(d-1)\right)^{d\over 2} V,}
$B_d$ being  the Bernoulli numbers.
With this choice the anomaly  for a conformal theory of a
 massless scalar is
\eqn\confanomaly{\l\theta\r_{cft}=  B_d (-R)^{d\over 2}=
 {A_d a^d\over V}\ ,}
thus,
\eqn\cscalar{c=1 \qquad {\rm for\ a\ massless\ scalar\ .}}
Furthermore, $c=11$ for free massless
fermions
and $c=62$ for vectors \refs{\duff,\birrell}.

The operator $\theta$ can be further decomposed as
\eqn\operatortheta{\theta= \theta_{an} + \theta_{dyn}\ ,}
where $\theta_{dyn}$, which we call
the {\sl dynamical} trace anomaly (also called
{\sl internal} anomaly), carries  all the dependence on the beta-functions
of the theory
\eqn\thetadynamical{\theta_{dyn}=\beta^i {\cal O}_i\ .}
Note that  masses are treated as couplings and therefore 
 contributions to the trace of the energy-momentum tensor
which are present at the classical level are included in $\theta_{dyn}$.
At fixed points, $\theta_{dyn}$ vanishes and 
 $\theta$ reduces to the background
anomaly, which is proportional to the identity operator:
\eqn\thetaanomalous{\l\theta\r_{cft}=\l\theta_{an}\r_{cft}=
c B_d (-R)^{d\over 2}\ .}

Because of the $x$--independence of  $\l\theta\r$,
the function $c({\mu\over a},g^i)$
 only depends on the subtraction point $\mu$ explicitly through
the dimensionless ratio
$\mu/ a$ and, implicitly, through the renormalized couplings $g^i$.
Furthermore, since the
energy--momentum tensor has vanishing
anomalous dimension (being a conserved current), the $c$--function
obeys the homogeneous RG equation
\eqn\cequation{\mu{\d\over \d\mu}c\left({\mu\over a},g^i\right)=
\left(\mu{\partial\over \partial\mu} +
\beta^i(g){\pr\over \pr g^i}\right) c\left({\mu\over a},g^i\right) =0\ .}
 Note that $g^i$ stands for all the couplings in the theory, including
those to the background, as well as masses and any other dimensionful 
parameters. The couplings are all rendered dimensionless by dividing
out the appropriate power of $\mu$ (as it is customary
in the Wilsonian renormalization group approach).
The explicit $\mu$ dependence 
can be traded for the dependence on the
only dimensionful physical parameter $a$, leading to\foot{
The same result can be of course obtained 
by first solving eq.~\cequation\ in terms of running 
couplings $\bar g^i\left({\Lambda\over a}\right)$, where $\Lambda$
is a RG invariant
dynamically generated scale,  and then differentiating 
with respect to $a$.}~\unpublished
\eqn\adepend{a{\d\over \d a}c\left({\mu\over a},g^i\right)=
\beta^j(g){\pr\over \pr g^j}c\left({\mu\over a},g^i\right) .}
Now, in a symmetric space, a rescaling of $a$ is equivalent to a general
scale transformation:
\eqn\scaltr{\delta_s\equiv a{\d\over \d a}= -{2}\int \d^dx\  g^{\alpha\beta}(x)
{\delta\over \delta g^{\alpha\beta}(x)}\ .}
However, the response of a generic Green function for a composite 
operator ${\cal O}(y)$ to scale
transformations is fixed by the scale Ward identity
\refs{\wirev\widet-\scwir}:
\eqn\genwi{-{2}g^{\alpha\beta}(x)
{\delta\over \delta g^{\alpha\beta}(x)} \langle{\cal O}(y)\rangle=-
\nabla_\mu\l j^\mu_D(x)
{\cal O}(y)\r+\delta^{(d)}(x-y)
\langle\delta_s{\cal O}(y)\rangle+\langle
\theta(x)  {\cal O}(y)\rangle\ ,}
where $j^\mu_D(x)$ is the dilatation current, whose charge generates
scale transformations.
The second term on the r.h.s. of eq.~\genwi\ 
gives the canonical scale transformation
of the operator $\delta_s{\cal O}(y)=\gamma_{\cal O}{\cal O}(y)$, where 
$\gamma_{\cal O}$ is
the dimension (engineering plus anomalous) of the operator ${\cal O}(y)$,
while the third term is the anomalous divergence of the dilatation
current, which satisfies the anomaly
equation
\eqn\opancons{\nabla_\mu j^\mu_D(x) =\theta(x)}
at the operator level. All the operators in eqs.~\genwi-\opancons\ are
renormalized composite operators~\widet; the expectation values
correspond to $T^*$--ordered proper Green functions,
i.e. only receive contributions from connected diagrams.\foot{It may
 be instructive to compare the scale Ward identity eq.~\genwi\ with
 the well--known chiral Ward identities which describe the way chiral
 symmetry is realized, for instance in QCD or in the sigma
 model. In the chiral case, the dilatation current $j^\mu_D$ is replaced
 by the axial current $j^\mu_5$. 
 Since chiral symmetry is an internal symmetry, the l.h.s. of
 eq.~\genwi, which corresponds to the transformation of the space-time
 coordinates, is then simply zero. In the absence of an axial anomaly (for
 instance in the non-singlet sector) the last term on the r.h.s. of
 eq.~\genwi\ also vanishes. The two remaining terms are then either
 separately zero, or else must cancel. The latter case, in which
 there exists a chirally non-invariant  operator $\cal O$ 
 with a non-vanishing vacuum expectation value, and a massless mode
 that couples to the divergence of the axial current, corresponds to the 
 Goldstone realization of the chiral symmetry.} 

The scale dependence of $c({\mu\over a},g^i)$
 can be obtained by integrating 
 the anomalous scale Ward identity eq.~\genwi\
over all space, and specializing to 
the case in which ${\cal
O}(x)=\theta(x)$.
Upon integration, the
contribution proportional to the divergence of the current
provides a surface term and thus vanishes exponentially
unless there is a massless particle (dilaton) in the spectrum
which couples to $\nabla_\mu j^\mu_D$ and $\theta$.
The anomalous dimension of $\theta$ vanishes
and its engineering
dimension coincides with the space dimension, so 
$\gamma_\theta=d$. Furthermore, the anomalous contribution $\theta_{an}$ to the
energy-momentum tensor eq.~\operatortheta, being  proportional to the
identity operator, can only contribute to the one--point Green function,
but not to $n$--point proper
Green functions of $\theta$.
It follows that the trace Ward identity for $\theta$ reads 
\eqn\thetwi{a{\d\over \d a} \langle\theta\rangle=
{1\over V}\int \d^dx\sqrt{g(x)}\langle
\theta_{dyn}(x)  \theta_{dyn}(0)\rangle_s+d\langle\theta\rangle,}
where the subscript on the two--point function indicates that 
the
surface (contact) contribution due to  the current divergence term
when dilatons are present 
has been included as a subtraction in the 
definition of the correlator. Note however that, quite  in general, no dilaton 
contribution is expected, in which case the correlators
$\l\theta\theta\r_s$ simply coincides with the $T^*$ ordered
correlator which appears in the Ward identity eq.~\genwi.

The scale transformation of $c({\mu\over a},g^i)$ 
is thus fully determined as
\eqn\changeofc{a{\d\over \d a} c = {a^{-d}\over A_d}
\int \d^dx \sqrt{g(x)} \l\theta_{dyn}(x) \theta_{dyn}(0)\r_s  \ .}
The contact term due to the canonical transformation of $\theta$ upon
dilatations is exactly canceled by an equal and opposite contact term
due to the dimensionful factor $a^{-d}$ which relates $\theta$ to the
dimensionless $c$-function.
Furthermore, 
\eqn\thth{
\l \left.\theta(x)\theta(0)\r\right|_{non\ local}\equiv
\l\theta(x)_{dyn}\theta(0)_{dyn}\r_s=
\beta^i\beta^j\l{\cal O}_i(x)
{\cal O}_j(0)\r_s\ \geq 0 ,}
where the last inequality can be derived from positivity of the spectral
representation along the lines of the
argument presented in sect.~2, and will be discussed in the next section.
We conclude that
\eqn\finalth{-\beta^i \pr_i c = -{a^{-d}\over A_d}\int \! \d^dx\ 
\sqrt{g(x)} \l\theta(x)_{dyn}\theta(0)_{dyn}\r_s \leq 0\ ,}
thus establishing the desired result, i.e. the monotonic decrease of
$c({\mu\over a},g^i)$ along RG flows.

Note that at fixed points $c$
coincides, up to a normalization, with the coefficient of the
Euler density in the trace anomaly and its derivative is
consistently zero.
Away from fixed points a positive definite, invertible, scheme-dependent
 metric in the space of couplings can  be defined as
\eqn\metric{G_{ij}={1\over V}\int
\!\d^d x\ \sqrt{g(x)} \l{\cal O}_i(x) {\cal
O}_j(0)\r_s \
.}
It is then easy to  show, along the lines of the two-dimensional
argument of ref.~\zamref, that to the first few orders
of conformal perturbation theory
\eqn\betas{\beta^i= G^{ij} \pr_j c\ ,}
i.e. the beta functions are gradients of the $c$--function.

The physical interpretation of eq. \finalth\ is simple. The
RG flow of $c({\mu\over a},g^i)$ is related by the Ward identity
to scale non-invariance and hence to the presence of massive
states in the Hilbert space.  The monotonic decrease of 
the $c$--function when moving towards the IR is
controlled by the decoupling
of these  massive modes
from the correlator eq.~\thth. 
At fixed
points, 
no such modes
are present and the $c$--charge is accordingly stationary. 
The $c$--function thus provides an effective counting of
massless degrees of freedom in the theory.
By deriving
the positivity of the correlator eq.~\thth\ from that of the spectral
representation (see sect.~5 below) the
monotonic character of the decoupling is directly derived from
the unitarity  of the spectrum.

We shall now discuss in somewhat greater detail some technical issues 
which we have set aside in our proof of the theorem, 
specifically in relation to
the choice of the background metric and to the precise definition of
subtractions in the definition of Green functions.

\bigskip
\goodbreak
\noindent{\sl Background metric}
\nobreak

As explained in sect.~3, the background metric is used as a device
to deal with the coefficient of the Euler density in the
trace anomaly, which is also present in correlators of
three point functions in flat space~\osborntwo.
 For the sake of definiteness, we
chose to present our argument in the hyperbolic space $H_d$. 
This space is maximally
symmetric and provides the natural Euclidean continuation of anti-de-Sitter
space. Unlike what happens on the sphere, massive
propagators decay at infinity, as we shall see in the
next section.
Nevertheless, the proof of the Ward identity
is easily generalized to a form which makes no
explicit reference to the choice of 
background metric.

Specifically, the $c$--function can be generally defined as
\eqn\ctwo{c_M=\int_M \!\d^d x\ \sqrt{g(x)} \l\theta(x)\r\ ,}
where $\theta$ is defined as the trace of the energy momentum tensor,
$\theta(x) \equiv T^\mu_\mu(x)$.
In general,
scale transformations are  implemented by \scaltr\ and it
 then follows that
\eqn\genericcth{
\delta_s c_M=
{-2\over V}\int \d^dx\ g^{\alpha\beta}(x){\delta\over
\delta g^{\alpha\beta}(x)}
\int \d^dy\ \sqrt{g(y)}\l\theta(y)\r.}
The contact term in the Ward identity eq.~\thetwi\ is then removed by
the contact term due to the variation of $\sqrt{g}$ on the r.h.s. of
eq.~\genericcth, thus implying 
\eqn\gencth{\delta_s c_M={1\over V} \int \d^dx \sqrt{g(x)}
\int \d^dy
\sqrt{g(y)}\ \l\theta_{dyn}(x) \theta_{dyn}(0)\r_s\ \ ,}
which is the general form of the Ward identity satisfied by the $c$--function.

Even though the Ward identity which expresses the scale variation of
the $c$--function is easily generalized to generic space, 
the proof of positivity of the correlator on the r.h.s. of eq.~\finalth\
in sect.~5 will make crucial use of the assumption
that space is maximally symmetric: on generic spaces, 
positivity of
the correlator~\gencth\ is not guaranteed.
In fact,
the $c$--charge eq.~\cardycandidate\ on a generic space will also pick up
contributions proportional to the $c_F$ and $c_R$ terms
 in eq.~\fandg, but it has been shown by explicit computation~\freedman\
that no combination of the $c_F$ and $c_G$ coefficients decreases 
in general along RG
flows.
Here, we are only interested in the validity of the $c$--theorem in the 
flat space limit. We will therefore
not address the issue of classifying the properties of the
background space which are necessary and sufficient for the $c$--theorem
to hold, and we will
limit ourselves to establishing our result
on $H_d$.

\bigskip
\goodbreak
\noindent{\sl Subtractions and scale Ward identities}
\nobreak

All the  arguments in the proof of irreversibility are done at the
level of renormalized correlators. This means that first, we must
define precisely the subtraction scheme used to define the one--point function
$\l\theta\r$, and furthermore, we must fix the contact terms which in
general  appear in  the definition of the two--point function
$\l\theta(x)\theta(y)\r$ .

The guiding principle in this  issue is provided
by the Ward identities.
Specifically, the scale Ward identity eq.~\genwi\ is derived upon the
assumption that the energy-momentum tensor is equal to the variation
of the renormalized action upon diffeomorphism, which in turns means
that
\eqn\defoft{\l \sqrt{g} T^{\alpha\beta}(x)\r = - 2 V
{\delta \ln Z\over\delta g_{\alpha\beta}(x)} }
(where $V$ is the
volume of the $d-1$--dimensional sphere eq.~\norm), 
so that the $\theta$ one--point function
is defined as
\eqn\defoftrt{
\sqrt{g(x)}\l\theta(x)\rangle=-2 V g_{\alpha\beta}(x)
{\delta\over \delta g_{\alpha\beta}(x)}\ln Z .}

Furthermore, the Ward identity eq.~\genwi\ only holds for the
specific
choice of contact terms in the definition of correlators of composite
operators such that
equations of motion are satisfied at the operator level by the 
renormalized operators \widet. For the correlator of two
traces of energy-momentum tensors this means that
\eqn\defoftt{\l 
\theta(x) \theta(y)\r
 =4 V^2
{g^{\mu\nu}(x)\over\sqrt{g}}{\delta \over\delta g_{\mu\nu}(x)}{
g^{\alpha\beta}(y)\over\sqrt{g}}
{\delta \over\delta g_{\alpha\beta}(y)}\ln Z
+2V\l{g^{\mu\nu}(x)\over\sqrt{g}}
{\delta\over\delta
 g_{\mu\nu}(x)} \theta(y)\r\ .}
With this definition, the correlator reduces to a purely dynamical
contribution and thus vanishes at conformal points.

Different definitions of the correlators are of course possible: for
 instance, in ref.~\cfl\ the correlator is constructed to satisfy
 covariant conservation
\eqn\conservation{\qquad
\nabla_\mu\l T^{\mu\nu}T^{\alpha\beta}\r_{CC}=0\ .}
This fixes the contact terms as 
\eqn\altdefoftt{\eqalign{\l \sqrt{g}
T^{\mu\nu}(x) \sqrt{g}T^{\alpha\beta}(y)\r_{CC}
=&4 V^2
{\delta \over\delta g_{\mu\nu}(x)}
{\delta \over\delta g_{\alpha\beta}(y)}\ln Z\cr&\qquad +
{V\over d} \langle
\theta\r \delta^d(x-y)\sqrt{g}
\left(g^{\mu\nu} g^{\alpha\beta}-
g^{\nu(\alpha} g^{\mu\beta)}\right)\ ,\cr}}
which then implies
\eqn\altwi{-2  g_{\alpha\beta}(x)
{\delta\over \delta g_{\alpha\beta}(x)}\langle\theta(y)\r={1\over V}
\sqrt{g(x)}\l\theta(x)\theta(y)\rangle_{CC}}
 instead of the Ward identity eq.~\genwi, i.e., with this definition, 
the contact term which
appears on the r.h.s. of the Ward identity is included in the
definition of the correlator.
It thus follows that 
\eqn\corrrelns{\l\theta(x)\theta(y)\rangle_{CC}=\l\theta(x)
\theta(y)\rangle_s+
d\ V \l
\theta\r{\delta^{(d)}(x-y)\over\sqrt{g}}.}
Of course,  the cancellation of contact terms leading to
eq.~\finalth, which follows from the fact that $c({\mu\over a},g^i)$ 
is a dimensionless
object, will always take place regardless of the choice of scheme;
however in different  schemes the Ward
identities will look different.

\bigskip\bigskip

\newsec{Spectral analysis and positivity of correlators}

Unitarity implies that all the states in the Hilbert space
of a theory
 have positive norm. The spectral representation of
a two--point correlator is based on saturating the correlator 
with a resolution of
the
identity made with all the physical states in the Hilbert
space of the theory. In this way, unitarity translates into positivity of
correlators. We will now use this general framework to prove the
positivity of the two--point function of $\theta_{dyn}$ eq.~\finalth\ 
which is needed to complete our proof.

The states in the  Hilbert space of a theory provide  representations of
the symmetry group of space-time. In our case, the background
geometry is the hyperboloid $H_d$,
which is understood as a Wick rotation of
$AdS_d$. The corresponding isometry group is then
$SO(d-1,2)$.
 A discussion of
the basic elements of this group and
the construction of spectral representations in $H_d$
can be found in refs.~\refs{\dusedau,\vilenki}\ and~\cfl\ whose notations and
conventions we will follow.

The spectral representation for the renormalized two--point
correlator of the trace of the energy-momentum tensor reads in general 
\eqn\spectralthth{
\l\theta(x)\theta(0)\r = A_d a^{d-2} \int _{\lambda_{min}}^\infty
\d\lambda\ \rho\left(\lambda;{\mu\over a},g^i\right)
 \left(\Delta+ d a^2\right)^2
G_\lambda(z)\ ,}
where 
$\lambda$ is the spectral parameter,
$\rho(\lambda;{\mu\over a},g^i)$
 is the positive spectral function which depends on
the spectral parameter,  on the dimensionless ratio $\mu / a$ and
on the renormalized couplings $g^i$,
and $G_\lambda$ is 
the propagator of a scalar particle carrying the
highest weight representation of $SO(d-1,2)$.
Notice that eq.~\spectralthth\ will hold for any choice of contact
 terms in the definition of the correlator.

The explicit form of the highest--weight propagator is
\eqn\propagatorhd{G_\lambda={1\over 2 \pi^{d\over 2}} \left(
{-a^2\over 2}\right)^{d-2\over 2} (z^2-1)^{2-d\over 4}
Q_{\lambda-{d\over 2}}^{d-2\over 2}(z)\ ,}
where $Q^\mu_\nu(z)$ are associated Legendre functions and
$z=\cosh a r$. 
This  propagator  decays at infinity as $z^{-\lambda}$ and 
satisfies by definition
\eqn\propeq{(\Delta+a^2  c_2(\lambda)) G_\lambda= {\delta^d(x)
\over \sqrt{g(x)}}\ ,}
where $c_2(\lambda)=\lambda (\lambda-d+1)$ is the second
Casimir operator of the group.
The unitarity bound for scalar representations is
$\lambda\geq (d-3)/2$. The value of $c_s(\lambda)$ is related to 
the mass of the intermediate state by
\eqn\mass{m^2-{d(d-2)\over 4} a^2 = a^2 \lambda(\lambda-d+1) \ . }
The second term in the l.h.s. of eq.~\mass\ is due to the coupling of
the fields to the curvature required to preserve conformal invariance
of the massless theory at the classical level;
 a massless scalar corresponds thus to the representation
$\lambda={d\over 2}$.

Using the definition of the propagator eq.~\propeq\ the spectral
representation eq.~\spectralthth\ can be explicitly rewritten as
\eqn\spectralththtwo{
\l\theta(x)\theta(0)\r = A_d a^{d-2} \int _{\lambda_{min}}^\infty 
\d\lambda\ \rho(\lambda;{\mu\over a}, g^i) \left(\Delta+ d a^2\right)^2
{1\over \Delta + a^2 \lambda(\lambda-d+1)} {\delta^d(x)\over \sqrt{g(x)}}
\ .}
{}From this expression it is apparent that, within the unitary region
$\lambda\ge(d-3)/2$, the spectral representation becomes local in
space if and only if $\lambda=d$. For this value, the correlator
reduces to a pure contact term:
\eqn\contactterm{\rho\left(\lambda;{\mu\over a},g^i\right)\propto 
\delta(\lambda-d)
\ \Rightarrow\ \l\theta(x)\theta(0)\r \propto  a^{d-2}
\left(\Delta+ d a^2\right){\delta^d(x)\over \sqrt{g(x)})} \ .}
This case corresponds to that of a
conformal field theory. This is thus also the lowest value of the spectral
parameter $\lambda$, i.e. $\lambda_{min}=d$, as required demanding
that
the correlator of two $c$--functions
 be well--behaved at spatial infinity~\cfl.

It is important to notice 
a fundamental difference between the spectral representation
in flat space (relevant for the two--dimensional construction of
sect.~2) and curved space. In flat space, the coefficient of the
delta contribution  is
constant on dimensional grounds. As a consequence, the IR limit is
non--uniform, and the IR and UV values of the central charge are
related by an additive scale--independent constant. 
In   curved space, an extra scale is available, so that the
general form of the local contribution to the spectral function is
\eqn\masslessmodes{\rho\left(\lambda;{\mu\over a}, g^i\right)=
\rho\left({\mu\over a},g^i\right)\ \delta(\lambda-d)\ .}
At fixed points the whole spectral function reduces to
\eqn\fixedpointspectral{
\rho\left({\mu\over a},g^i\right)
\longrightarrow \rho_{cft}\qquad\Rightarrow\qquad
\rho(\lambda,t)\longrightarrow \rho_{cft}\ \delta(\lambda-d)
\ .}
However the value $\rho_{cft}$ is attained through smooth variation of the
function $\rho\left(\lambda;{\mu\over a},g^i\right)$ along the flow;
therefore, we cannot simply derive the irreversibility of the flow
from a
comparison of the UV and IR limits, as in sect.~2: a study of the
derivative of the $c$--function along the flow, such as that of
sect.~4, is required to prove
the $c$--theorem.

Now, in order to prove positivity of the correlator which gives the
evolution of our $c$--function, eq.~\finalth, we must specify the
precise definition of the correlator. The definition used in the Ward
identity is characterized by the fact that only $\theta_{dyn}$,
proportional to the beta--functions, contributes to it.
Furthermore, the surface (contact) contribution coming from the 
current divergence term upon integration of the Ward identity eq.~\genwi\
 is 
also included in the correlator. However,
consider surface contributions to
the unintegrated Ward identity eq.~\genwi\ with
${\cal O}=c\left({\mu\over a},g^i\right)$.
 The l.h.s. of the equation is
manifestly vanishing, since surface contributions to $c({\mu\over a},g^i)$ 
single out its constant IR
limit. Therefore, on the r.h.s. 
the surface contribution from the current divergence must exactly
cancel the current correlator. But at infinity massive contributions
decouple, so the latter also
reduces to the
contribution of intermediate massless states to the spectral
decomposition of eq.~\spectralththtwo. In other words, the 
contribution of intermediate  massless states to the $\theta$--correlator
exactly matches the surface contribution
from the divergence of
the dilatation current, consistently with the anomalous conservation law
eq.~\opancons. The physical origin of the ensuing cancellation is
simple: the anomalous scale dependence of the $c$--function is driven
by the scale--dependence of the vacuum, controlled by the trace
anomaly through the $\l \theta\theta\r_s$ correlator, and  massless
intermediate states decouple from it.

The correlator appearing in the
Ward identity \thetwi\ is thus free of contact terms:
\eqn\ambiguityfree{
\l\theta_{dyn}(x) \theta_{dyn}(0)\r_s =
A_d a^{d-2} \int _{\lambda=d+\epsilon}^\infty
\d\lambda\ \rho\left(\lambda;{\mu\over a},g^i\right) 
\left(\Delta+ d a^2\right)^2
G_\lambda(z)\ ,}
where $\epsilon$ denotes that
the contact term has been  removed. All intermediate states
are massive and yield non-local contributions
to the correlator. At fixed points the whole correlator
vanishes since massless modes do not contribute to it.
Then, upon integration we find
\eqn\spesumrule{
\int \d^d x \sqrt{g(x)} \l\theta_{dyn}(x)\theta_{dyn}(0)\r =
{A_d d^2\over 2^{d-2\over 2}} a^d
\int_{\lambda=d+\epsilon}^\infty 
\!\d\lambda\ {\rho\left(\lambda;{\mu\over a}, g^i\right)\over
\lambda(\lambda-d+1)}\geq 0\ ,}
which is a scheme--independent statement and establishes the result we
set out to prove:
unitarity imposes the positivity for the correlator that
controls the sign of the change of $c({\mu\over a},g^i)$.
 Notice that the explicit
form of the spectral representation eq.~\spectralththtwo, the
cancellation of long--distance contribution responsible for the
absence of contact terms in the correlator eq.~\ambiguityfree, and the
positivity of the spectral function $\rho$ are all derived upon the 
assumption that space is maximally symmetric.

\bigskip\bigskip

\newsec{An explicit example}

In order to illustrate our proof of the $c$--theorem we study now a
 simple explicit example, namely, the case of a free massive
scalar
field in the hyperbolic background 
$H_4$.
We will thus be able to check  
the scale Ward identity and the
manifest positivity of the $c$--function in a simple setting.

The relevant field theory is discussed in 
refs.~\refs{\camporesi,\cfl,\burgess} to which we refer  for details.
The bare trace of
the energy-momentum tensor is 
\eqn\baretrace{\theta(x) = V m^2 \phi^2(x)\ ,}
with
\eqn\masses{{m^2\over a^2}=(\lambda-2)(\lambda-1)=\sigma^2
-{1\over 4} \ .}
We can  verify explicitly our $c$--theorem eq.~\finalth\ by
computing the $c$--function, demonstrating that its derivative is indeed
given by a correlator free of contact terms, and checking the
positivity of this correlator.

The $c$--function is given by eq.~\candidate\ in terms of the trace
anomaly which, for the theory at hand, has been determined in
ref.~\camporesi\ (see also ref.~\burgess):
\eqn\traceanomaly{a^{-4} \l\theta\r =
30 \left( \left({m\over a}\right)^4 \left(
\psi\left(\sigma+{1\over 2}\right) - \log {\nu\over a}\right) -
{3\over 4} \left({m\over a}\right)^4 -{1\over 6} \left({m\over a}\right)^2
+{1\over 30}\right)\ ,}
where $\psi(z)$ is the polygamma function and 
$\nu$ is a renormalization scale. In order to fix the renormalization
scheme, we require that 
the $m\to \infty$ limit  (IR limit) be a
trivial fixed point. This corresponds to the fact that in this limit
there are no  propagating modes. This means that we require
$\l\theta\r_{IR}\equiv \lim_{m\to\infty}\l\theta\r =0$, which fixes 
uniquely $\log {m\over \nu}={3\over 4}$. The
$c$--function is thus completely determined as
\eqn\scalarcfun{
c\left({m\over a}\right)= 30 \left(
\left({m\over a}\right)^4 \left( \psi\left(\sigma+{1\over 2}\right)
 - \log {m
\over a}\right) -{1\over 6} \left({m\over a}\right)^2 + {1\over 30}
\right) .}

In order to verify the theorem we must now compute 
the correlator of two energy-momentum tensors, which, recalling that
the scalar propagator is $G_\lambda$, is simply found to be
\eqn\twott{\l \theta(x)\theta(0)\r_{pert} = 2 V^2 m^4
\left(G_\lambda(z)\right)^2 \ . }
This expression manifestly needs renormalization since the 
product of two propagators which appears on the r.h.s. is not a
distribution.
Furthermore, the perturbative definition of the correlator implicit
in eq.~\twott\ 
does not necessarily coincide to that used in the Ward
identity~\genwi, so
we will also
need to perform an extra subtraction
in order to be able to compare to eq.~\finalth.

The UV divergence of the correlator is clearly seen when
considering the integrated correlator,
\eqn\intcorr{
\int\d^4x\ \sqrt{g(x)} \l\theta(x)\theta(0)\r_{pert} =
 {V^3\over 8\pi^4} m^4 \int_1^{\infty} \d z\ \left(Q_{\lambda-2}^1(z)
\right)^2\ ,}
which displays a log singularity at $z=1$. A simple way to 
renormalize it consists of using differential renormalization~\dr,
whereby  the integrand on the r.h.s. of eq.~\intcorr\ is
rewritten by pulling out a derivative, which is then
understood to act on its left side in the sense of distributions.
This softens the singularity and renders it integrable while
leaving the integrand unchanged away from the singular point. In our case,
this amounts to writing
\eqn\renampl{
\left(Q_{\lambda-2}^1(z)\right)^2 =
-{\darrow{\d}\over \d z}
\left(\sqrt{z^2-1} Q_{\lambda-2}(z) Q_{\lambda-2}^1(z)\right) -
 Q_{\lambda-2}(z) {\d\over \d z}
\left(\sqrt{z^2-1} Q_{\lambda-2}^1(z)\right)\ ,}
where the total derivative is understood to act on its left.
We can then perform the $z$-integral and get a finite result
by using
\eqn\intstep{\int_1^\infty \d z \left(Q_{\lambda-2}(z)\right)^2=
{\psi'(\lambda-{1\over 2})\over \lambda-3}\ .}

This result is still only defined up to a choice of renormalization
scheme, i.e. a polynomial in $m$, 
which we may again fix by requiring the IR limit to be
trivial, i.e. by subtracting from the correlator its $m\to\infty$
limit. This prescription in principle still allows the subtraction of 
terms which are subleading in the IR
limit, i.e. inverse powers of $m$. These must however all vanish
because they would diverge in the
UV limit. 
Note also that no massless infrared modes are present in this
example and thus no subtraction of the surface term related to these
modes (as discussed
below eq. \thetwi) is required.
We thus get
\eqn\finalthth{\int \d^4x\ \sqrt{g(x)}\ \l\theta(x)\theta(0)\r_{pert}=
-{\pi^2\over 2} a^4 \left( \left({m\over a}\right)^6
{\psi'(\sigma+{1\over 2})\over \sigma }
- \left({m\over a}\right)^4 +{1\over 3} \left({m\over a}\right)^2
-{2\over 15}\right)\ .}

We can now determine the subtraction required to define the correlator
\eqn\subtrc{\int \d^4x\ \sqrt{g(x)}\ \l\theta(x)\theta(0)\r=
\int \d^4x\ \sqrt{g(x)}\ \l\theta(x)\theta(0)\r_{pert}+\kappa\
c\left({m\over a}\right)}
by imposing that the correlator 
vanishes at the conformal point,
i.e., specifically, that
it vanishes as $m\to 0$. This fixes  $\kappa=-4$, thus showing
that the perturbative definition of the correlator coincides with the
covariantly conserved one eq.~\altdefoftt.
With this choice we see explicitly that the correlator is equal to
the derivative of the $c$--function, thus verifying the $c$--theorem
eq.~\finalth:
\eqn\check{\eqalign{
a{\pr\over \pr a} c\left({m\over a}\right) =&
30 \left[ -\left({m\over a}\right)^6 {\psi'(\sigma+{1\over 2})\over
\sigma}  - 4 \left({m\over a}\right)^4 \psi\left(\sigma+{1\over 2}\right)
 \right.\cr
 &\left. +
4 \left({m\over a}\right)^4 \log{m\over a} + \left({m\over a}\right)^4
+{1\over 3} \left({m\over a}\right)^2\right]\ .\cr }}

We can now check explicitly that
the $c$--function and its derivative satisfy the desired
properties.
Note in particular that according to eq.~\check\ the derivative of the
$c$--function is positive
and proportional to $m^2$, that is, to the square of the beta--function.
Plots of the $c$--function and its derivative are displayed in Fig.~1.
We can finally get the
Zamolodchikov metric directly from
eq.~\check, by dividing it by $m^2$: the metric is
positive definite and invertible away from the fixed points.

\topinsert
\vbox{
\hfil\epsfysize=4.5cm\epsfbox{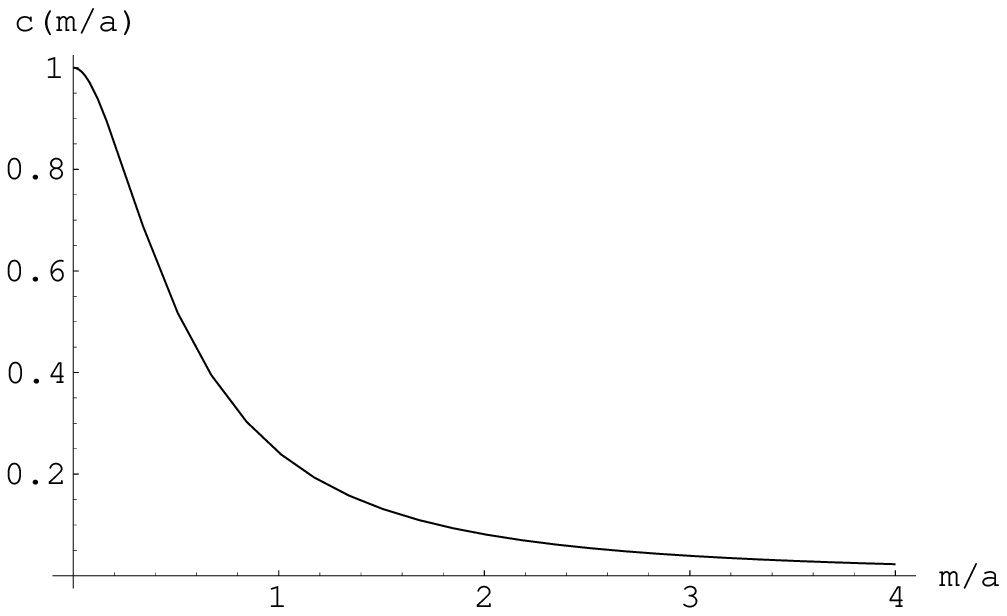}\hfil}
\vbox{
\hfil\epsfysize=4.5cm\epsfbox{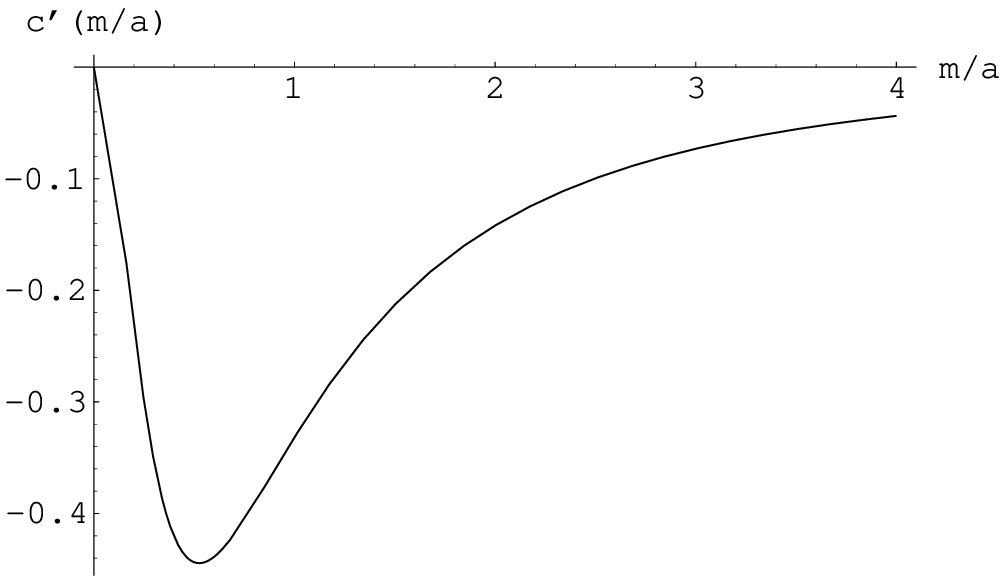}\hfil}
\bigskip\noindent{\footnotefont\baselineskip6pt\narrower
Figure 1: The function $c(m/a)$ and its logarithmic derivative
$c'(m/a)\equiv -a{d\over da}c(m/a) $.
\medskip}
\endinsert

\bigskip\bigskip

\newsec{Anomaly constraints}

We can summarize  the results we have obtained in the following way.
Irreversibility of RG flows in 
four--dimensional
(or, more generally, even-dimensional) renormalizable quantum field
theories follows from unitarity.  
The integrated trace anomaly plays the role of a $c$--function which
decreases  along RG trajectories. This $c$--function is
stationary at fixed points, where it gives a $c$--charge that coincides
with the coefficient of the Euler density in the trace anomaly.
Irreversibility is established using
the scale Ward identities, and affords a 
very natural physical interpretation: the decrease of the 
$c$--function  along the RG flow is due to the decoupling of 
intermediate massive
states from the Hilbert space. At fixed points the 
$c$--charge is additive and 
counts effective degrees of freedom.

The $c$--theorem provides a new instrument to relate short-- {\sl vs.}
long--distance realizations of quantum field theories. The set of constraints
emerging from
the combination of 't Hooft anomaly matching and the irreversibility of
RG flows can be dubbed {\sl anomaly constraints}. Axial and trace
anomalies are characterized at fixed points
by coefficients (charges)
 that multiply topological terms. The behavior of these charges under RG 
transformations is not the same: axial anomalies are protected by the 
Adler-Bardeen non-renormalization theorem and thus scale--independent,
 whereas the trace anomaly
scales in an irreversible way. 

The power of anomaly matching relies in that it requires exact
matching
of the charge at all scales, while its drawback
comes from the need of the corresponding axial
symmetry to be present in the
theory in the first place. On the other hand, the decrease 
of the $c$--charge along
RG flows is of more general nature since it exists for any theory. 
The constraint coming from its irreversibility, however, 
is an inequality rather than an exact equality.

The way this inequality is satisfied seems to be related to the
way symmetries are realized at various scales. Indeed, with the
definition \cardy\ of the $c$--function which we studied,
the $c$--charge  of free fermions and vectors are larger than
the one associated to scalars. This suggests that, very crudely
speaking, scalars should be ``preferred" in  IR realizations. This 
agrees with the empirical 
observation that in many cases in which the decrease of
the $c$--function can 
be verified by explicit computation, this turns out to be due to the fact
 that the IR theory displays
 spontaneous chiral symmetry breaking and thus contains 
Nambu-Goldstone scalar bosons. One is thus lead to speculate that 
the fact that chiral symmetry is realized in the Nambu-Goldstone mode
at low energy and in the Wigner-Weyl mode at high energy in realistic
theories such as QCD may be due to a deep interplay between 
scale and chiral symmetry.

\bigskip
\goodbreak
\noindent{\bf Acknowledgments}
\nobreak
We thank D. Z. Freedman, H. Osborn and R. Tarrach for 
a critical reading of the manuscript.
S.F. is supported by an IBERDROLA visiting professorship at the
University of Barcelona; J.I.L. acknowledges financial support from 
CICYT (contract AEN95-0590),
CIRIT (contract 1996GR00066) and 
NATO (contract CRG 910890).

\listrefs
\bye